\documentclass{PoS}

\title{Inclination Dependence of The Time-Lag -- Photon-Index Correlation in BHXRBs and its Explanation with a Simple Jet Model}

\ShortTitle{Time-Lag -- Photon-Index Correlation in BHXRBs}

\author{\speaker{N. Kylafis}\thanks{A footnote may follow.}\\
        Physics Department, University of Crete, 70013 Heraklion, Greece\\
	Institute of astrophysics, FORTH, 71110 Heraklion, Greece\\
        E-mail: \email{kylafis@physics.uoc.gr}}

\author{P. Reig\\
        Institute of astrophysics, FORTH, 71110 Heraklion, Greece\\
	Physics Department, University of Crete, 70013 Heraklion, Greece\\
        E-mail: \email{pau@physics.uoc.gr}}

\abstract{
Recently, we reported an observational correlation between a) the time-lag
of the hard (9 - 15 keV) with respect to the soft (2 - 5 keV) X-ray
photons in black-hole X-ray binaries (BHXRBs) and 
b) the power-law photon index $\Gamma$ of the X-ray spectrum.
This was physically explained with a simple jet model, i.e., a model where
the Comptonization (the Compton upscattering of soft photons) happens 
in the jet.  Here, we report the inclination dependence of this correlation,
which we also explain with our jet model.  Photons that emerge at different
polar angles from the jet axis have different spectra and different time-lags.
Because of this, we can explain {{\it quantitatively}} 
the type-B QPOs of GX 339-4 as resulting from a precessing jet.
}

\FullConference{High Energy Phenomena in Relativistic Outflows VII - HEPRO VII\\
		9-12 July 2019\\
		Facultat de F\'{\i}sica, Universitat de Barcelona, Spain}

\begin{document}

\section{Introduction}

Before entering into the details of this presentation, we would like to 
comment on two strong beliefs that our community has.

The first strong belief is that the power-law X-ray spectrum in black-hole 
X-ray Binaries (BHXRBs) is produced in the, so called, ``corona" around the 
black hole \cite{done07} or at the base of the jet \cite{markoff05}.
Since the ``corona" cannot be static, it is natural to take it to be the
inner hot part of the accretion flow, a region that is ADAF-like
\cite{narayan94,narayan95}.  
Outside this region, the accretion flow is in the form of a
Shakura-Sunyaev disk (SS disk, \cite{shakura73}).  
The picture is then that soft photons from the SS disk get upscattered
by the electrons in the hot inner flow or the base of the jet and thus
the power-law spectrum with index $\Gamma$ is produced.  There is nothing
wrong with this picture, but it may be incomplete.

The second strong belief is that the average time-lag of the hard X-ray photons
(say, 9 - 15 keV) with resect to softer ones (say, 2 - 5 keV) is caused by
propagating fluctuations in the accretion flow \cite{kotov01,arevalo06}.  
There is nothing wrong with this either, and it is nearly certain that it 
happens in accretion flows, but the two
mechanisms (Comptonization and propagating fluctuations) ``do not talk" to 
each other.  In other words, the two mechanisms seem to work independently,
and no correlation is expected between 
the time-lag (of hard photons with respect to softer ones)
and the photon number spectral index $\Gamma$ in the observed power law.
However, these two quantities are correlated!

\section{Recent developments}

A nice correlation (Fig. 1) has been found in the BHXRB GX 339-4
\cite{kylafis18}.  One can see that, as the source moves from the quiescent
state to the hard state and then to the hard intermediate one (filled circles),
$\Gamma$ increases monotonically and the time-lag increases in the hard state
and then flattens out.

%------------------------------Fig. 1-----------------------------
\begin{figure}
\centering
\includegraphics[angle=0,width=8cm]{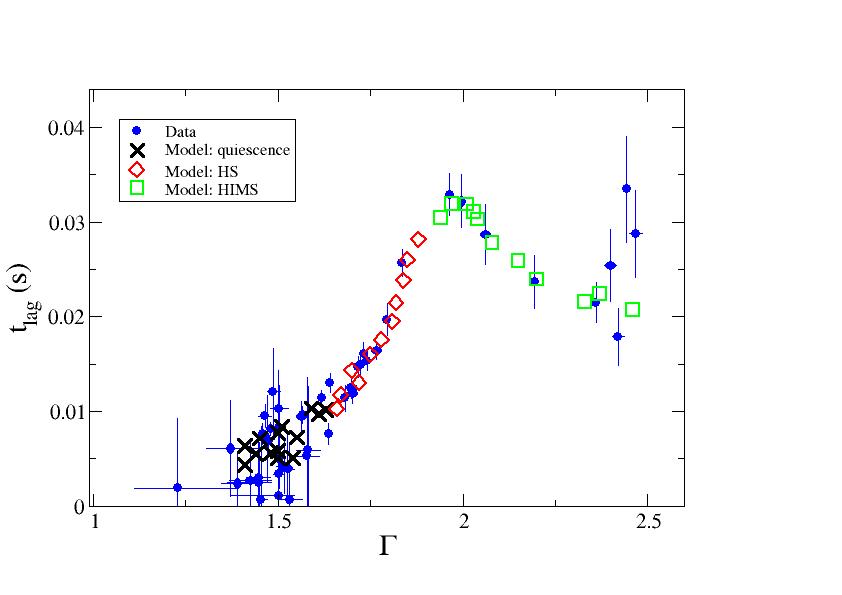}
\caption{Observed correlation between the time-lag and $\Gamma$ for
GX 339-4}
\label{t_lag-Gamma: GX 339-4}
\end{figure}
%-----------------------------------------------------------------

We have explained this correlation with a simple jet model
\cite{kylafis18}.  The model is the same, as the one we used
before, to explain the energy spectra \cite{giannios05}, the
dependence of the time-lags on Fourier frequency \cite{reig03}, 
the correlation between the time-lag and $\Gamma$ in Cyg X--1 
(Kylafis et al. 2008), and the correlation between
time-lag and cut-off energy in GX 339--4 \cite{reig15}.

\section{The jet model}

The basic idea of the jet model is that soft photons from the SS disk 
get upscattered in the jet, which is taken to be parabolic, as the 
observations suggest 
\cite{asada12}.  For a quantitative description
of the jet model the reader is referred to \cite{reig19}, and references therein.

The physical picture that we have in mind is the following:  the accretion 
flow consists of two parts, the outer SS disk and the inner
hot flow, which is called ``corona".  Soft photons from the SS disk 
either escape and are detected as such or they are upscattered in the 
``corona".  Some of these upscattered photons escape, but most of them
enter the jet, which lies above the ``corona" and it is fed by it. There,
they are scattered again and this final Comptonization produces
the observed hard X-ray power law with photon index $\Gamma$.

We want to stress here that Comptonization in the jet is unavoidable,
because most of the photons that enter the ``corona" must go through
the jet.  Since photons ``forget" their
past history after a few scatterings, it is the scatterings in the jet
that leave their imprint on the observed spectra and time-lags.

On average, the more times a photon is scattered in the jet, the larger
its energy becomes and the more is delayed in escaping from the jet, 
due to light-travel time.
Since the same mechanism produces both the high-energy power law
and the time-lags, it is natural that the two quantities are correlated.

It is important to remark that, if the soft photons from the SS disk
enter in the base of the jet,
nothing can prevent them from exporing the whole jet.  And indeed, this
is what happens.

\section{New developments}

One may wonder if the correlation shown in Fig. 1 is just a peculiarity  of GX
339-4.  The answer is no.  In Fig. 2, we show the time-lag as a function of
$\Gamma$ for 13 sources, for which we could find data for our analysis
\cite{reig18}.  The straight line shows the correlation that exists between the
time-lag and $\Gamma$. The correlation is mathematically acceptable, i.e. the
slope of the regression is not consistent with zero (see \cite{reig18} for
details of the statistical analysis), but it is not pleasing, because there is a
lot of scatter.

%------------------------------Fig. 2-----------------------------
\begin{figure}
\centering
\includegraphics[angle=0,width=8cm]{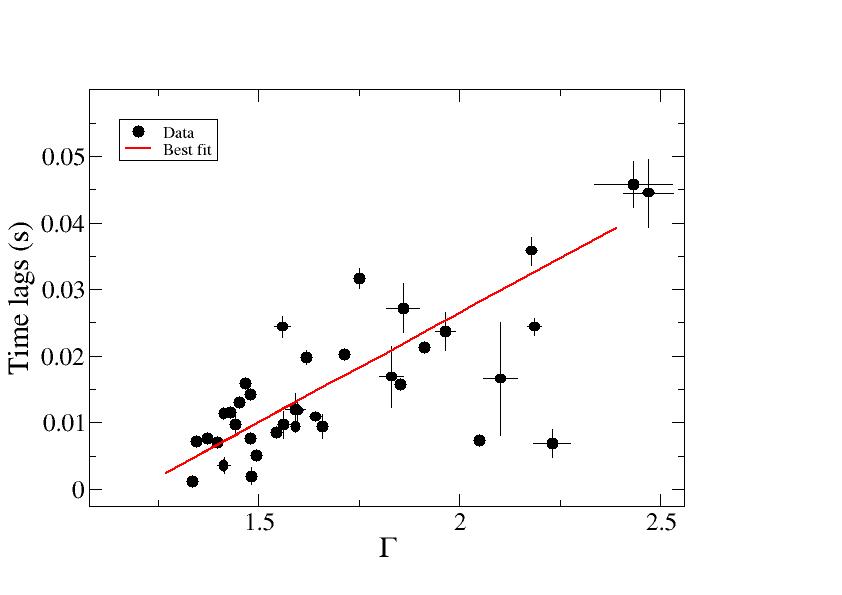}
\caption{Observed correlation between the time-lag and $\Gamma$ for
13 sources.}
\label{t_lag-Gamma: 13 sources}
\end{figure}
%-----------------------------------------------------------------

We examined whether this scatter is due to the different inclination  of the
sources.  Indeed, this is the case.  In Fig. 3 we show the time-lag versus
$\Gamma$ correlation for low-inclination sources (sources for which the
accretion flow is seen nearly face-on, filled circles) and for high-inclination
ones (sources for which the accretion flow is seen  approximately edge-on, empty
circles. It is clear from Fig. 3, that the scatter seen in Fig. 2 can be easily
accounted for by the inclination of the sources.

%------------------------------Fig. 3 -----------------------------
\begin{figure}
\centering
\includegraphics[angle=0,width=8cm]{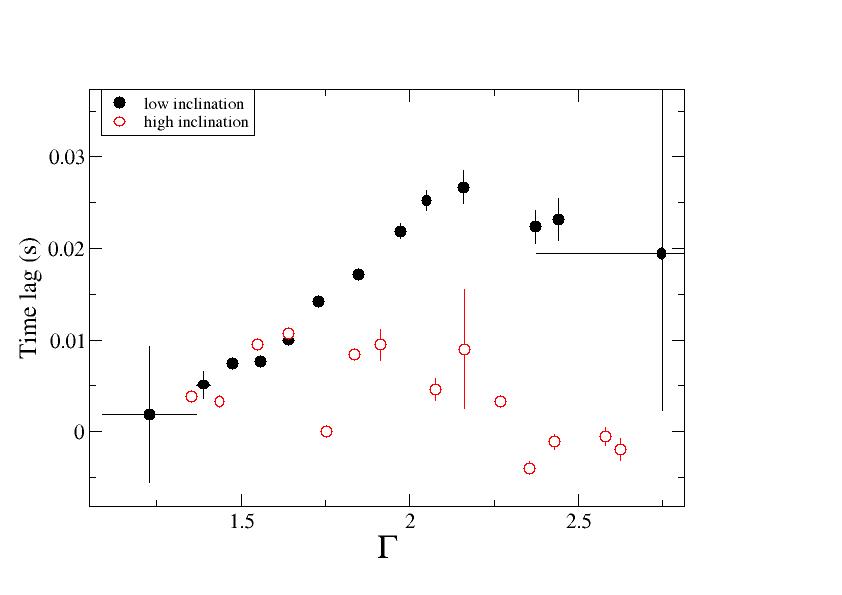}
\caption{Observed correlation between the time-lag and $\Gamma$ for 
low-inclination sources (filled circles) and for high-inclination ones
(empty circles).  
}
\label{t_lag-Gamma: high-low}
\end{figure}
%-----------------------------------------------------------------

This led us to examine in more detail the inclination dependence of the
time-lag versus $\Gamma$ correlation in BHXRBs.  We considered 17 sources
and took into account all the data that we could find.  In order to have
relativey good statistics, we divided the sources into three categories:
low-nclination ($0 \le \theta < 30$ degrees), intermediate inclination
($30 \le \theta < 70$ degrees), and high inclination
($70 \le \theta \le 90$ degrees) \cite{reig19}.

The results are shown in Fig. 4.  The filled symbols (squares, circles, and
triangles) represent the data, while the stars represent the 
results of our jet model calculations. For these model calculations, we have
varied two parameters, the optical depth to Thomson scattering along
the jet and the radius of the jet at its base.  For details, the 
reader is referred to \cite{reig19}.

The model  parameters, optical depth and radius of the jet have been chosen so
that a good fit is made to the data of the top panel, i.e., the low-inclination
sources.   Then, using the {\it same model parameters}, we derived  the
jet-model predictions for intermediate observing angles ($30 \le \theta < 70$
degrees, stars in the middle panel of Fig. 4) and large observing angles ($70
\le \theta \le 90$ degrees, stars in the lower panel of Fig. 4). The resemblance
of the model results to the real ones is rather impressive.

%--------------------------------Fig. 4--------------------------------
\begin{figure}
\centering
\includegraphics[angle=0,width=9cm]{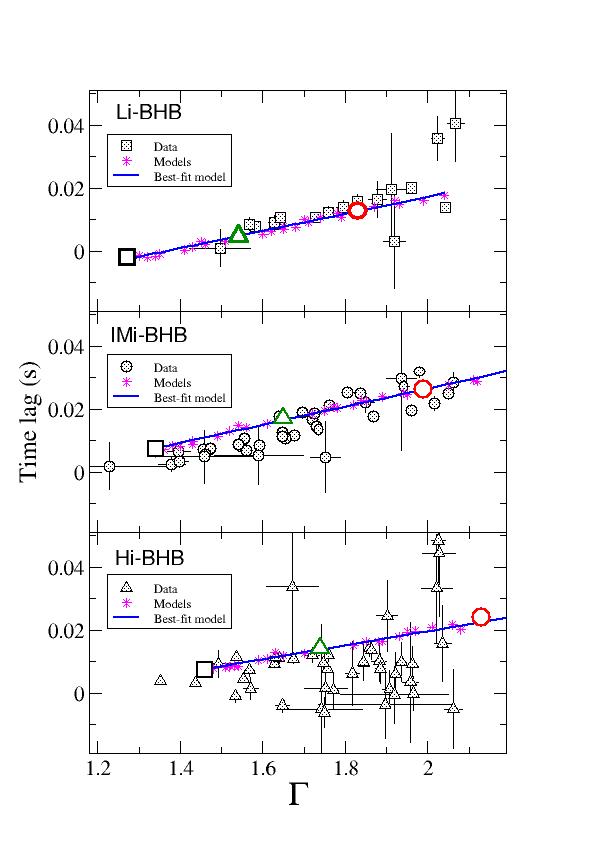}
\caption{Correlation between the time-lag and the photon index for individual
systems. {\em Top panel:} low-inclination systems. {\em Middle panel:}
intermediate-inclination systems. {\em Bottom panel:} high-inclination systems.
Different sources are represented by different symbols.}
\label{indiv}
\end{figure}
%----------------------------------------------------------------------

Of all the jet model calculations, that resulted in the star symbols  in Fig. 4,
we pay special attention to three model calculations, indicated by an empty
square, an empty triangle and an empty circle.  It is clear that, if we could
see the same source from different directions, both the  time-lag and $\Gamma$
would be different.  In particular, the spectra would become softer (i.e.,
larger $\Gamma$) as the observational angle $\theta$ increases from 0 to 90
degrees.  This is understood by the fact that the bulk motion in the jet
produces a harder spectrum along the jet than perpendicular to it.

One may say that this is practically irrelevant, because we cannot see
the same source from different directions.  Not so, if the source of
hard X-ray photons is the jet and the jet is precessing!  In such a case,
we would see a periodic variation of $\Gamma$ with period the precession
period.

We repeat here, that in the model calculations
used in Fig. 4, we varied only two 
parameters:  the optical depth to Thomson scattering along the jet
$\tau_{\parallel}$ and the radius of the jet at its base $R_0$.  Any 
worries, that two parameters are too many, are alleviated by the fact
that the two parameters are correlated!  Fig. 5 shows the values of
$\tau_{\parallel}$ and $R_0$ that have been used 
and they are nicely correlated.  So, the two
parameters are really one.

%------------------------------Fig. 5-------------------------------
\begin{figure}
\centering
\includegraphics[angle=0,width=8cm]{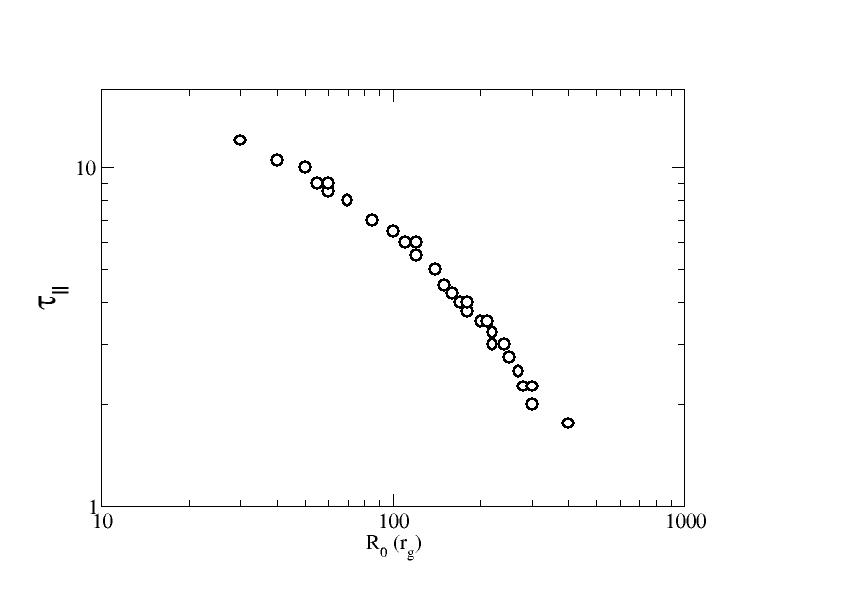}
\caption{Relationship between the optical depth
along the jet $\tau_{\parallel}$ and the
radius $R_0$ at the base of the jet, for the models that reproduce the 
correlations.}
\label{tau-R0}
\end{figure}
%-------------------------------------------------------------------

We further remark that it is difficult for ``corona" models, i.e., models 
where the Comptonization takes place in the ``corona", to explain the
inclination dependence of $\Gamma$, let alone the correlation of
time-lag versus $\Gamma$.
This is because in the ``corona" the electrons are thermal and therefore
scattering is isotropic.  Thus, no inclination dependence is expected.
In a jet, however, the electrons have a
bulk velocity, which makes the scattered spectra anisotropic.  A harder
spectrum is produced in the forward direction than perpendicular to it.

\section{B-type QPOs}

Phase-resolved spectroscopy of the type-B  Quasi Periodic Oscillations (QPOs) in
GX\,339-4 was performed by \cite{stevens16} and found a sinusoidal
variation of $\Gamma$ with phase.  They interpreted it as a precessing jet.
This is exactly what we found above!  A precessing jet should exhibit
a sinusoidal variation of $\Gamma$ with phase.

To quantify this variation, we calculated with our jet model  and specific
parameters the variation of $\Gamma$ with viewing angle $\theta$.  In the left
panel of Fig. 6, we  show model results with the same $\tau_{\parallel}$ and
various $R_0$ and in the  right panel model results with the same $R_0$ and
various $\tau_{\parallel}$. The horizontal dotted lines mark the range of the
variation of $\Gamma$ that was found by \cite{stevens16}.  The vertical dotted
lines bracket the cosine of the inclination angle $\theta$, which is thought to
be 45 degrees \cite{motta18}.  

%------------------------------Fig. 6 -------------------------------------
\begin{figure}
\centering
\includegraphics[angle=0,width=8cm]{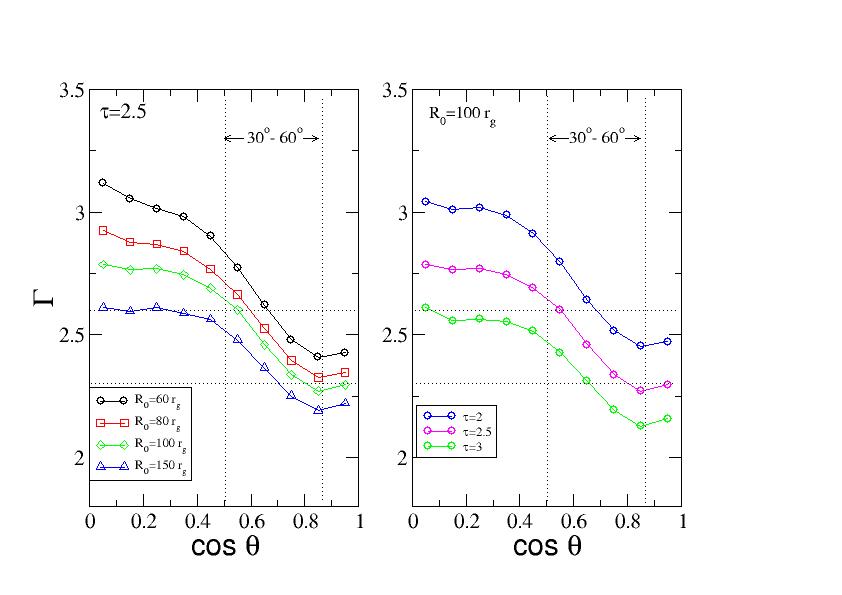}
\caption{Model results for the relation between $\Gamma$ and the polar angle
$\theta$ of the direction of observation.
Left panel: models with fixed $\tau_{\parallel}$ and various values of $R_0$.
Right panel: models with fixed $R_0$ and various $\tau_{\parallel}$.}
\label{Gamma-cos(theta)}
\end{figure}
%--------------------------------------------------------------------------

Notice that the constraints placed by the observations are quite stringent
and both the left and the right panels select models with 
$\tau_{\parallel} \approx 2.5$ and $R_0 \approx 100 R_g$, where $R_g$ is
the gravirational radius of a 10 solar-mass black hole.

\section{Conclusions}

Comptonization in the jet is {\it unavoidable} and it is very important,
because it is the last one before the photons escape.  Comptonization in
the ``corona" is possible, but it is irrelevant if it is followed by
Comptonization in the jet.

Even if one has a favorable mechanism for the time-lags, e.g.
propagating fluctuations, one must also take into account the lags that 
are necessarily introduced by the Compton scattering in the ``corona".  
In other words, {\it ``corona" models should specify the size and shape of the 
``corona"}.

Last, we comment on the so called lamp-post model for studying the 
reflection from the accretion disk.  In this model, a point source
is placed in the rotation axis of the accretion disk at a height above 
the black hole \cite{ross05,garcia10}.  
Recently, \cite{garcia19} found
that, in order to explain the reflected spectrum
in GX 339-4, they had to introduce
two point sources, one at a few $R_g$ and one at $500 R_g$.

The lamp-post model is a nice mathematical tool for studying reflection,
but it is totally unphysical.  On the other hand, the jet constitutes
a ``natural lamp post"!  In an upcoming publication
(Reig \& Kylafis, in preparation), we will show the fraction of
downward scattered photons, i.e., from the jet to the accretion disk,
as functions of energy and height of emission.

\end{document}